\documentclass{elsart}

 \def\ket{\!>\,} \def\ack{\,|\,}

\usepackage{amssymb}
\usepackage{graphicx}

\begin{document}

\begin{frontmatter}

\title{Multi-phonon $\gamma$-vibrational bands in odd-mass nuclei studied
by triaxial projected shell model approach}

\author[a1,a2]{J. A. Sheikh,}
\author[a1]{G. H. Bhat,}
\author[a3,a4,a2]{Y. Sun,}
\author[a5]{R. Palit}

\address[a1]{Department of Physics, University of Kashmir, Srinagar
190 006, India}
\address[a2]{Department of Physics and Astronomy, University of
Tennessee, Knoxville, TN 37996, USA}
\address[a3]{Department of Physics, Shanghai Jiao Tong University,
Shanghai 200240, P. R. China}
\address[a4]{Institute of Modern Physics, Chinese Academy of Sciences,
Lanzhou 730000, P. R. China}
\address[a5]{Department of Nuclear and Atomic Physics,
Tata Institute of Fundamental Research, Colaba, Mumbai, India}

\begin{abstract}

Inspired by the recent experimental data (Phys. Lett. B {\bf 675}
(2009) 420), we extend the triaxial projected shell model approach
to study the $\gamma$-band structure in odd-mass nuclei. As a first
application of the new development, the $\gamma$-vibrational
structure of $^{103}$Nb is investigated.  It is demonstrated that
the model describes the ground-state band and multi-phonon
$\gamma$-vibrations quite satisfactorily, supporting the
interpretation of the data as one of the few experimentally-known
examples of simultaneous occurrence of one- and two-$\gamma$-phonon
vibrational bands.  This generalizes the well-known concept of the
surface $\gamma$-oscillation in deformed nuclei built on the
ground-state in even-even systems to $\gamma$-bands based on
quasiparticle configurations in odd-mass systems.

\end{abstract}

\end{frontmatter}

Atomic nucleus is a many-body quantal system exhibiting pronounced
shell effects, which give rise to intrinsic deformation.  In
addition, it can, according to the semiclassical collective model,
undergo dynamical oscillations around the equilibrium shape,
resulting in various low-lying collective excitations.  Ellipsoidal
oscillation of the shape is commonly termed $\gamma$-vibration
\cite{BM75}. Rotational bands based on the $\gamma$-vibrational
states are known as $\gamma$-bands.  The interplay between
rotational and vibrational degrees of freedom plays a central role
in our understanding of structure of atomic nuclei.  One-phonon
$\gamma$-bands have been well known and observed in numerous
deformed nuclei in most of the regions of the periodic table.
However, observation of two or higher order phonon $\gamma$-bands is
a rare event possibly because, due to excitation, these bands are
embedded in the energy region with dense levels.  Nevertheless,
there have been reports on successful observation of two-phonon
$\gamma$-band ($\gamma\gamma$-band) in well-deformed even-even
nuclei \cite{Fah96,Har98}.  Recently, fission experiment data have
suggested simultaneous observation of one-phonon $\gamma$- and
two-phonon $\gamma\gamma$-bands in odd-mass systems
\cite{Ding06,Gu09a,Gu09b,Wang09}.

While the physics of one-phonon $\gamma$-bands seems to be well
understood, there has been a considerable debate on the nature of
$\gamma\gamma$-bands.  The issue is whether these bands are really
two-phonon excitations built on the rotational states or they are
based on an ensemble of two-quasiparticle excitations.  Theoretical
investigations using quasiparticle-phonon nuclear model (QPNM)
\cite{SS81,Sol92} predicted that $\gamma\gamma$-vibrational bands
cannot exist in deformed nuclei due to Pauli blocking of
quasiparticle components.  On the other hand, the multi-phonon
method (MPM) \cite{LP88,JP88} suggested that a two-phonon $K^\pi$ =
4$^+$ state (in even-even nuclei) should appear at an excitation
energy of about 2.6 times the energy of the one-phonon $K^\pi$ =
2$^+$ state, and the decay from the $\gamma\gamma$- to $\gamma$-band
should be predominantly collective in character. Strictly speaking,
due to the violation of rotational symmetry, these methods do not
calculate the states of angular-momentum, but the $K$ states (where
$K$ is the projection of angular-momentum on the intrinsic symmetry
axis). Therefore, the QPNM and MPM models do not have their wave
functions as eigenstates of angular-momentum, and consequently, the
reliability of these predictions depends critically on the actual
situation.  As pointed out by Soloviev \cite{Sol92}, it is quite
desirable to recover good angular-momentum in the wave function.

Recently, the triaxial projected shell model (TPSM) approach has
been developed and applied to investigate $\gamma$-bands in
transitional even-even nuclei \cite{Sun00,Sun02,Sheikh08,Sheikh09}.
The TPSM approach \cite{Sheikh99}, generalized form the
one-dimensional projected shell model \cite{PSM}, performs explicit
three-dimensional angular-momentum projection from triaxially
deformed Nilsson states.  Shell model diagonalization is carried out
in the laboratory frame with good angular-momentum, thus removing
uncertainties caused by angular-momentum non-conservation in the
QPNM and MPM models.

In the TPSM, each triaxially-deformed configuration is an admixture
of different $K$ states.  For example, the quasiparticle vacuum
configuration is composed of $K=0,2,4,\cdots$ states for an
even-even system.  It has been demonstrated in Ref. \cite{Sun00}
that the angular-momentum-projection on these $K=0,2$ and 4 vacuum
states generates the low-spin parts of the ground, $\gamma$-, and
$\gamma\gamma$-band, respectively.  This is a pleasant feature of
the TPSM that the quantum-mechanical treatment of
angular-momentum-projection on the (triaxially deformed)
quasiparticle vacuum alone gives rise to rotational ground band and
multiphonon $\gamma$-vibrational bands.  Of course, this simple
configuration \cite{Sun00} is valid only for low-spin states because
if the system goes to high spins, rotational alignment will bring
various quasiparticle configurations down to the yrast region, and
therefore, the quasiparticle excitations must be considered.  In our
recent work \cite{Sheikh08,Sheikh09}, the TPSM has been generalized
to include two-neutron and two-proton quasiparticle states for
even-even nuclei.  It has been shown that indeed, $\gamma$-bands
built on these quasiparticle states can become favoured at higher
angular-momenta.  Using this generalization, we have offered an
explanation for the long-standing puzzle of the observation of two
aligned $I^\pi=10^+$ states with same intrinsic neutron structure in
the mass-130 region.  It has been demonstrated \cite{Sheikh09} that
$\gamma$-band built on the neutron aligned configuration is the
first excited band above the parent aligned configuration, and this,
together with the standard neutron aligned state itself, provides a
natural explanation of the observation of two aligned structures
with same intrinsic configuration.

In a very recent experimental paper \cite{Wang09}, $\gamma$- and
$\gamma\gamma$-bands have been reported in an odd-proton system,
$^{103}$Nb, by measuring prompt $\gamma$-rays following the
spontaneous fission of $^{252}$Cf.  The ground state of this nucleus
has the spin-parity $K^\pi_g=5/2^+$.  Based on this, the
experimentally observed $\gamma$-band built on $9/2^+$ and
$\gamma\gamma$-band based on $13/2^+$ have been proposed.  Naively,
the spin-2 phonon structure in an odd-mass nucleus is quite
analogous to that in an even-even system based on the $0^+$ ground
state and one can have the $\gamma$-band built on $K_\gamma=K_g+2$
and $\gamma\gamma$-band based on $K_{\gamma\gamma}=K_g+4$. However,
there has been, so far, no theoretical work that describes this
observation, and thus supporting the interpretation.  The purpose of
this Letter is to study these band structures using the newly
developed TPSM approach.

The present study generalizes, for the first time, the TPSM to
odd-mass nuclei with the inclusion of quasiparticle (qp)
configurations in the model basis.  For the study of odd-proton
system, our model space is spanned by (angular-momentum-projected)
one- and three-qp basis
\begin{equation}
\{ \hat P^I_{MK}~a^\dagger_{p} \ack\Phi\ket , ~\hat
P^I_{MK}~a^\dagger_{p} a^\dagger_{n1} a^\dagger_{n2}\ack\Phi\ket \},
\label{basis}
\end{equation}
where the projector
\begin{equation}
\hat P^I_{MK} = {2I+1 \over 8\pi^2} \int d\Omega\,
D^{I}_{MK}(\Omega)\, \hat R(\Omega),
\end{equation}
and $\ack\Phi\ket$ represents the triaxially-deformed qp vacuum
state.  The qp basis chosen in (\ref{basis}) includes the
configurations of two-neutron aligned states built on the
one-quasiproton states.  The basis, with one- and three-qp
configurations included, has proven adequate to describe the
high-spin states in odd-mass systems and the rotation alignment
process \cite{Hara92}.  For odd-proton nuclei in this mass region,
it has been suggested \cite{Hua02} that the aligning neutrons are
from the $h_{11/2}$ orbital which is included in the calculation
(see below).  It should be noted that in the present case of
triaxial deformation, any qp-state is a superposition of all
possible $K$-values.  The rotational bands with the triaxial basis
states (\ref{basis}) are obtained by specifying different values for
the $K$-quantum number in the projection operator, Eq. (2). The
allowed values of the $K$-quantum number for a given intrinsic state
are obtained through the following symmetry requirement.  For $\hat
S = e^{-\imath \pi \hat J_z}$, we have
\begin{equation}
\hat P^I_{MK}\ack\Phi\ket = P^I_{MK} \hat S^{\dagger} \hat S \ack\Phi\ket
                          = e^{\imath \pi (K-\kappa)} P^I_{MK}\ack\Phi\ket,
                          \label{condition}
\end{equation}
where $ \hat S\ack\Phi\ket = e^{-\imath \pi \kappa}\ack\Phi\ket$. It
is easy to see that for the self-conjugate vacuum or 0-qp state,
$\kappa=0$ and, therefore, it follows from Eq. (\ref{condition})
that only $K=$ even values are permitted for this state.  For two-qp
states, the possible values for $K$-quantum number are both even and
odd, depending on the structure of the qp state.  For one-qp state,
$\kappa=1/2$ $(-1/2)$, and the possible values of $K$ are therefore
$1/2,5/2,9/2,\dots$ ($3/2,7/2,11/2,\dots$) that satisfy
Eq.~(\ref{condition}).

As in the earlier PSM calculations, we use the quadrupole-quadrupole
plus pairing Hamiltonian \cite{PSM}
\begin{equation}
\hat H = \hat H_0 - {1 \over 2} \chi \sum_\mu \hat Q^\dagger_\mu
\hat Q^{}_\mu - G_M \hat P^\dagger \hat P - G_Q \sum_\mu \hat
P^\dagger_\mu\hat P^{}_\mu . \label{hamham}
\end{equation}
The corresponding triaxial Nilsson mean-field Hamiltonian is given
by
\begin{equation}
\hat H_N = \hat H_0 - {2 \over 3}\hbar\omega\left\{\epsilon\hat Q_0
+\epsilon'{{\hat Q_{+2}+\hat Q_{-2}}\over\sqrt{2}}\right\},
\label{nilsson}
\end{equation}
where $\epsilon$ and $\epsilon'$ specify the axial and triaxial
deformations, respectively.  In the above equations, $\hat H_0$ is
the spherical single-particle Hamiltonian, which contains a proper
spin-orbit force.  The interaction strengths are taken as follows:
The $QQ$-force strength $\chi$ in Eq. (\ref{hamham}) is adjusted
such that the physical quadrupole deformation $\epsilon$ is obtained
as a result of the self-consistent mean-field calculation
\cite{PSM}.  The monopole pairing strength $G_M$ is of the standard
form: $G_M = \left[20.25 \mp16.20(N-Z)/A\right]/A$, with ``$-$" for
neutrons and ``$+$" for protons.  This choice of $G_M$ is
appropriate for the single-particle space employed in the present
calculation, where three major oscillation shells are used for each
type of nucleons ($N=3,4,5$ for neutrons and $N=2,3,4$ for protons).
The quadrupole pairing strength $G_Q$ is assumed to be proportional
to $G_M$, the proportionality constant being fixed as usual to be
0.16 \cite{PSM}.  These interaction strengths are, although not
exactly the same, consistent with those used earlier in the PSM
calculations.

\begin{figure}
\includegraphics[totalheight=10.5cm]{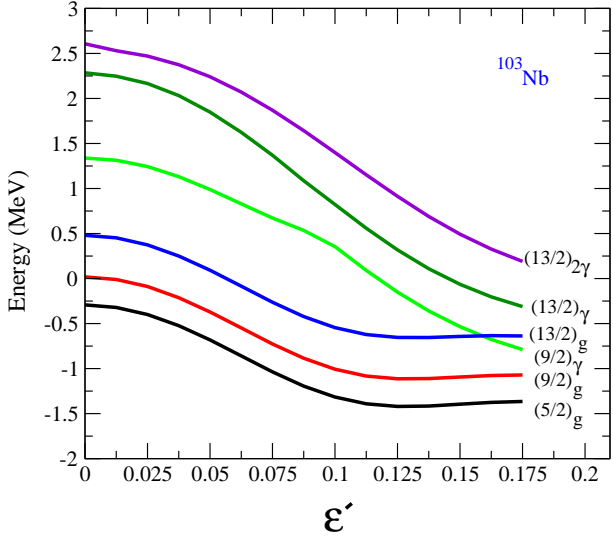}
\caption{(Color online) Variation of the projected energy surfaces
as functions of triaxiality $\epsilon'$. Shown in the plot are three
members of the $K=5/2$ ground band (denoted with a sub-index ``g''), two
members of the $K=9/2$ $\gamma$-band (denoted with a sub-index
$\gamma$), and one state of the $K=13/2$ $\gamma\gamma$-band
(denoted with a sub-index 2$\gamma$).} \label{fig1}
\end{figure}

Calculations have been performed for $^{103}$Nb with the quadrupole
deformation parameter $\epsilon=0.3$ and the triaxial one
$\epsilon'=0.16$.  The value of $\epsilon$ has been chosen from the
total-routhian-surface (TRS) calculations based on the cranked shell
model approach with Woods-Saxon potential and Strutinisky shell
correction formalism.  This value has also been employed in the
analysis of the experimental data in Ref.~\cite{Wang09}.  Nuclei in
this mass region are known to exhibit triaxiality \cite{Luo05}.  The
value of $\epsilon'$ in our calculation has been chosen such that
the bandhead energy of the experimental $\gamma$-band is reproduced,
which will be seen later.  In some of our earlier studies for
triaxially deformed nuclei, $\epsilon'$ was fixed from the
minimization of the ground state energy as a function of this
parameter (see, for example, Ref. \cite{Sheikh09}).  For the present
case, we have also calculated ground state energy as a function of
$\epsilon'$, as shown in Fig. \ref{fig1}, where states belonging to
the ground band are denoted with a sub-index ``g'' and those to
$\gamma$-band with a sub-index $\gamma$.  As one can clearly see
from the figure, the energy curves of the states of the ground band
are flat in the region of $\epsilon'=$ 0.12 - 0.17, indicating that
these states are very soft with respect to triaxiality.  In
contrast, energies of the states of the $\gamma$-band depend
sensitively on $\epsilon'$, showing low excitation energies for
large triaxiality.  This picture thus suggests that the nucleus does
not have a well-defined triaxial deformation in its ground state
configuration.  The same picture has been noted in some even-even
nuclear systems \cite{Sun00,Sun08}.  Therefore, $\epsilon'$ in the
present treatment is a chosen parameter that reproduces the
experimental bandhead of the $\gamma$-band and is equal to 0.16. It
is pertinent to mention that for the axially symmetric case
($\epsilon'=0$), $\gamma$-bandhead energy is calculated to be 1.63
MeV above the ground-state and the corresponding experimental values
is 0.73 MeV.  $\epsilon'=0.16$ leads to the $\gamma$-bandhead energy
of 0.72 MeV, which almost reproduces the experimental bandhead
energy.

\begin{figure}
\includegraphics[totalheight=11cm]{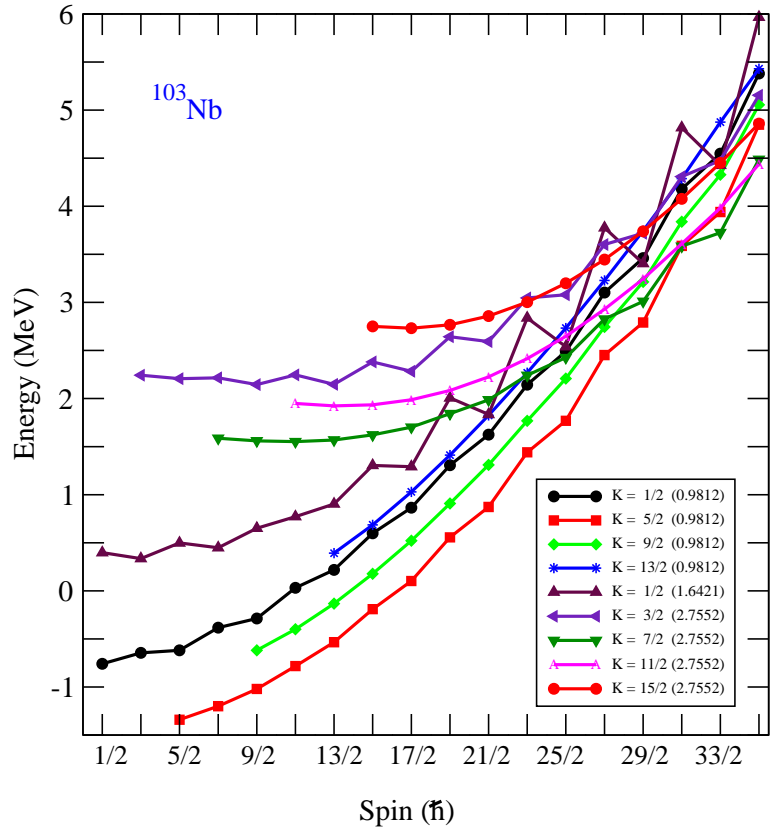}
\caption{(Color online) Band diagram for $^{103}$Nb depicting the
angular-momentum projected bands from one- and three-quasiparticle
states. For clarity, only the lowest projected bands are shown and
in the numerical calculations, projection has been performed from
forty-four intrinsic states.} \label{fig2}
\end{figure}

The TPSM analysis is performed in two stages: In the first stage, a
set of quasiparticle states are chosen for which the
angular-momentum-projection is to be performed.  In the second
stage, the shell model Hamiltonian is diagonalized in the projected
states. In the present work, we have considered an energy window of
3.5 MeV around the Fermi surfaces for both protons and neutrons, and
qp states of (\ref{basis}) with energy smaller than the window value
will be selected to form the model basis.  As mentioned earlier, the
possible values for one-qp state are $K=1/2, 5/2, 9/2, \cdots$ and
the projection from all these $K$-states has been performed.  In the
present calculation it is found that for protons, the lowest one-qp
state has an energy of 0.9812 MeV and projection from $K=5/2$ of
this configuration is the main component of the ground state band.
We note that this band depicts a staggering for higher
angular-momentum states.  This lowest band and the other low-lying
excited bands are shown in the band diagram, Fig. \ref{fig2}.

The projection from $K=9/2$ with the same lowest one-qp state is the
main component of the $\gamma$-band built on the ground state of
$K=5/2$, and the projection with $K=13/2$ gives rise to the
$\gamma\gamma$-band.  The calculated $\gamma$- and
$\gamma\gamma$-bandheads are at excitation energies of 0.73 and 1.73
MeV, respectively.  These bandhead energies will be modified by
diagonalization, i.e., configuration mixing.  It is noted from Fig.
\ref{fig2} that projection from the lowest qp state with $K=1/2$
lies higher in energy than the $\gamma$-band with $K=9/2$.  In
general, in odd-A nuclei, there exist two one-phonon $\gamma$-bands
with $K_\gamma=K\pm 2$, where $K$ is the bandhead quantum number of
the quasiparticle state.  The excitation energy of $K-2$ state is
larger than that of $K+2$ state \cite{piep90,Wang09}, and in the
present analysis corresponds to the $K=1/2$ band.

The projection from the qp state with an energy of 1.642 MeV for
$K=1/2$ is also shown in Fig. \ref{fig2} as this is a low-lying
band.  We have also carried out projection from this qp state for
other possible values of $K$, but they are not shown as these lie at
higher excitation energy.  It needs to be pointed out that all these
states, though not shown here, are employed in the diaganolization
of the shell model Hamiltonian.

For three-qp states, one-proton qp state can be coupled with
two-neutron states which have $\kappa=0$ and 1 and, therefore,
three-qp states can also have all possible $K$-values.  In Fig.
\ref{fig2}, only those three-qp bands with lower qp energies are
plotted, although the projection has been performed for all other
possible $K$-values.  In the present calculation, the lowest
three-qp intrinsic state has an excitation energy of 2.755 MeV, and
the projection from this three-qp state with $K=7/2$ forms the
lowest three-qp band.  This three-qp band is the two-neutron aligned
band built on the one-qp band having $K=5/2$.  It has already been
pointed out in our earlier publications \cite{Sheikh08,Sheikh09}
that each qp state has $\gamma$-bands built on it.  The $\gamma$-
and $\gamma\gamma$-bands built on the three-qp state with energy of
2.755 MeV have $K=11/2$ and 15/2, respectively, which are also
depicted in Fig. \ref{fig2}.  It is evident from the figure that the
three-qp band crosses the ground state band at $I=31/2$ and above
this spin value, the yrast band has mainly a three-qp structure.  It
is quite interesting to note from Fig. \ref{fig2} that $\gamma$- and
$\gamma\gamma$-band built on the three-qp band also cross the
corresponding one-qp bands.

\begin{figure}[t]
\includegraphics[totalheight=8.5cm]{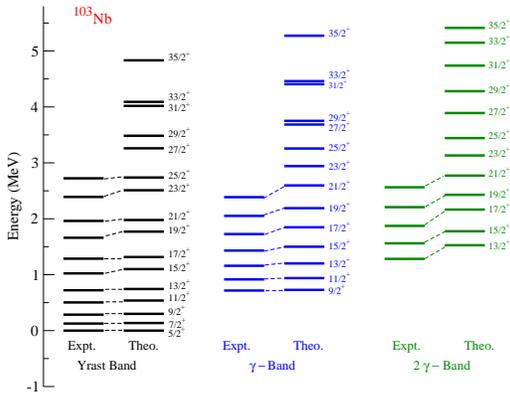}
\caption{(Color online) The calculated yrast band, $\gamma$- and
$\gamma\gamma$-bands are compared with the corresponding
experimental data \protect\cite{Wang09}.} \label{fig3}
\end{figure}

In the second stage of the present calculation, the projected states
obtained from a total of forty-four intrinsic states are used as
basis for diagonalizing the shell model Hamiltonian (\ref{hamham}).
The final energy levels after diagonalization are displayed in Fig.
\ref{fig3} for the yrast-, $\gamma$- and 2$\gamma$-bands.  This
figure also depicts the corresponding experimental band structures
obtained in Ref. \cite{Wang09}.  It is quite evident from Fig.
\ref{fig3} that TPSM describes the yrast- and $\gamma$-bands
remarkably well.  For the $\gamma\gamma$-band, the TPSM predicted
band lies higher by about 200 keV as compared to the experimental
$\gamma\gamma$-band.

The experimental data for $^{103}$Nb points towards anharmonic
$\gamma$-vibration with the ratio of the bandhead energies of
$\gamma\gamma$- and $\gamma$-band,
$E_{\gamma\gamma}/E_{\gamma}=1.87$.  The TPSM on the other hand
predicts a harmonic $\gamma$-vibration with
$E_{\gamma\gamma}/E_{\gamma}=2.01$.  Of course, the TPSM prediction
depends on the choice of the input parameter $\epsilon'$, which can
be clearly seen from Fig. \ref{fig1}.  A smaller $\epsilon'$ than
the used value of 0.16 will produce anharmonicity.  However, a
smaller $\epsilon'$ worsens the achieved agreement of the
$\gamma$-band.  The reason for this difficulty could be traced to
the $\gamma$-softnees of this nucleus, evident from the
energy-surface calculations in Fig. \ref{fig1}.  In a more accurate
treatment, the projection needs to be performed for different values
of $\epsilon'$ and then admix these projected states using the
generator coordinated method (GCM). We expect that this kind of
analysis may possibly explain the anharmonic $\gamma$-vibration
observed in $^{103}$Nb.

In conclusion, recent experimental data obtained by the spontaneous
fission of $^{252}$Cf have found $\gamma$- and $\gamma\gamma$-bands
in the odd-proton nucleus $^{103}$Nb.  Motivated by this, we have
extended the traixial projected shell model approach to odd-mass
systems with the inclusion of three-dimensional projected one- and
three-qp configurations in the shell model space.  As a first
application, the band structures of $^{103}$Nb have been
investigated.  It has been demonstrated that the observed yrast- and
$\gamma$-bands are reproduced quite well by the TPSM approach, thus
supporting the interpretation of the data.  However, for the
$\gamma\gamma$-band, there appears an overall shift in the energies,
resulting a more harmonic phonon vibration picture than suggested by
data.  We have proposed that this discrepancy could possibly be
resolved by performing GCM calculations with triaxial deformation
$\epsilon'$ as the generator coordinate.

Valuable discussions with S.-J. Zhu are acknowledged, which
stimulated our interest in the present work.  Y.S. was supported in
part by the National Natural Science Foundation of China under
contract No. 10875077, the Special Program of Chinese High Education
Foundation through grant 20090073110061, and the Chinese Major State
Basic Research Development Program through grant 2007CB815005.

\end{document}